\documentclass[twoside,journey]{IEEEtran}

\usepackage{makecell}
\usepackage{hyperref}
\usepackage{array}
\usepackage{caption2}
\usepackage{graphicx,amssymb,amsmath}
\usepackage{multicol}
\usepackage[noadjust]{cite}
\usepackage{setspace}
\usepackage{subfigure}
\usepackage{graphicx}
\usepackage{float}
\usepackage{url}
\usepackage{stfloats}
\usepackage{amsthm,pifont}
\usepackage{flushend}
\usepackage{cases,subeqnarray}
\usepackage{bm,multirow,bigstrut}
\usepackage{amsmath, amsthm, amssymb}
\usepackage{textcomp}
\usepackage{latexsym,bm}
\usepackage{booktabs}
\usepackage{xcolor}
\usepackage{mathtools}
\usepackage{dsfont}
\usepackage{extarrows}
\usepackage{epsfig}
\usepackage{epsfig}
\usepackage{epstopdf}
\usepackage[noend]{algpseudocode}
\usepackage{algorithmicx,algorithm}

\theoremstyle{plain}

\theoremstyle{plain}

\usepackage{amsmath}

\IEEEoverridecommandlockouts
\begin{document}

\title{Context-Aware Semantic Communication for the Wireless Networks}

\author{Guangyuan Liu, Yinqiu Liu, Jiacheng Wang, Hongyang Du, Dusit Niyato,~\IEEEmembership{Fellow,~IEEE}, Jiawen~Kang, \\Zehui~Xiong, and Abbas Jamalipour,~\IEEEmembership{Fellow,~IEEE}
\thanks{G.~Liu is with the College of Computing and Data Science, the Energy Research Institute @ NTU, Interdisciplinary Graduate Program, Nanyang Technological University, Singapore (e-mail: liug0022@e.ntu.edu.sg).}
\thanks{Y.~Liu, J.~Wang, D. Niyato are with the College of Computing and Data Science, Nanyang Technological University, Singapore (e-mail: yinqiu001@e.ntu.edu.sg, jiacheng.wang@ntu.edu.sg, dniyato@ntu.edu.sg).}
\thanks{H.~Du is with the Department of Electrical and Electronic Engineering, University of Hong Kong, Hong Kong (e-mail: duhy@eee.hku.hk).}
\thanks{J. Kang is with the School of Automation, Guangdong University of Technology, China. (e-mail: kavinkang@gdut.edu.cn).}
\thanks{Z. Xiong is with the Pillar of Information Systems Technology and Design, Singapore University of Technology and Design, Singapore (e-mail: zehui\_xiong@sutd.edu.sg).}
\thanks{A. Jamalipour is with the School of Electrical and Information Engineering, University of Sydney, Australia (e-mail: a.jamalipour@ieee.org).}}
\maketitle

\begin{abstract}

In next-generation wireless networks, supporting real-time applications such as augmented reality, autonomous driving, and immersive Metaverse services demands stringent constraints on bandwidth, latency, and reliability. Existing semantic communication (SemCom) approaches typically rely on static models, overlooking dynamic conditions and contextual cues vital for efficient transmission. To address these challenges, we propose CaSemCom, a context-aware SemCom framework that leverages a Large Language Model (LLM)–based gating mechanism and a Mixture of Experts (MoE) architecture to adaptively select and encode only high-impact semantic features across multiple data modalities. Our multimodal, multi-user case study demonstrates that CaSemCom significantly improves reconstructed image fidelity while reducing bandwidth usage, outperforming single-agent deep reinforcement learning (DRL) methods and traditional baselines in convergence speed, semantic accuracy, and retransmission overhead.
\end{abstract}


\section{Introduction} \label{sec:intro}

The fast growth of data-intensive applications, such as immersive Metaverse experiences, autonomous vehicles, and real-time language translation, has accelerated the quest for more efficient communication strategies in wireless networks~\cite{yang2022semantic}. With the advent of 5G/6G technologies, wireless systems are expected to meet stringent requirements for ultra-low latency, high bandwidth, and massive connectivity~\cite{wood2008context}. Based on Shannon's theory, traditional bit-level transmission focuses on the fidelity of bit recovery while paying little attention to the meaning of the data. Consequently, network resources are frequently expended on the transmission of extensive volumes of information that may not be pertinent to specific tasks or user requirements. Semantic communication (SemCom) has emerged as a transformative paradigm, aiming to encode and transmit only information relevant to the user’s objective. By doing so, it substantially reduces bandwidth usage and latency while preserving the intended meaning~\cite{yang2022semantic}.

However, existing SemCom schemes often face two significant challenges when deployed in practical wireless networks. Firstly, SemCom typically relies on a single static model or method to extract semantic features regardless of whether the data is text, voice, or video. Existing SemCom framework may not dynamically adapt to the specific requirements of the task, channel quality, and device constraints. Secondly, existing SemCom methods lacks robust context awareness. The context in modern heterogeneous wireless networks which range from millimeter-wave 5G macro cells to low-power Internet of Things (IoT) nodes can be multifaceted: includes user intentions (e.g., low latency for augmented reality), communication context (e.g., channel state information and interference levels), and device context (e.g., battery status, computational resources)~\cite{wood2008context}. Without effective context integration, SemCom systems may extract and transmit irrelevant or redundant semantic information, wasting resources and degrading performance. For instance, in a vehicle collision avoidance scenario, without considering its context, the system might focus on irrelevant details, such as cars in adjacent lanes, instead of prioritizing the critical semantic information—such as the nearest pedestrian in the vehicle's path. This misalignment can lead to unnecessary overhead, excessive delay, and critical safety conditions when the network quality deteriorates.

To address these challenges, we propose a context-aware SemCom (CaSemCom) framework, designed specifically for wireless networks. The core innovation is a Large Language Model (LLM)–based agentic context gate that orchestrates both input content selection and expert selection. The framework continuously collects task context, for example, whether the application is VR streaming or real-time object detection and communication context, such as channel state information or available bandwidth. Based on these inputs, CaSemCom first identifies the portions of raw data that carry the most critical semantic information, and then selects the specialized semantic-extraction expert model that should process the data. This multi-expert approach draws inspiration from Mixture of Experts (MoE) paradigms, and leverages an LLM to interpret a wide range of contextual cues, enabling more precise and adaptive gating decisions~\cite{du2024mixture}. As a result, the system can flexibly respond to fluctuations in wireless channel conditions, varying user demands, and heterogeneous device capabilities. By transmitting only the most relevant information, it achieves high semantic fidelity\footnote{Semantic fidelity refers to the preservation of the meaning or intent of the transmitted data, ensuring that the receiver accurately understands the intended semantics, even when the original data is compressed or altered for efficiency.} at a fraction of the bandwidth cost required by conventional methods. The contributions of this work can be summarized as follows:
\begin{itemize}
    \item \textbf{Context-aware SemCom:} We systemically discuss the role of context in improving SemCom's efficiency. Specifically, we analyze the types of context and their functionalities in the linguistic, wireless, and semantic communication paradigms. 
    \item \textbf{CaSemCom Framework:} We develop a novel context-aware SemCom framework named CaSemCom. Particularly, CaSeCom leverages an LLM-based agentic gate for context cognition and semantic/expert selection. Additionally, different semantic encoders are organized by an MoE architecture, ensuring efficiency and resilience under varying task and wireless channel conditions.  
    {
    \item \textbf{Case Study:} We present a comprehensive multimodal, multi-user case study under realistic Rayleigh-fading channel conditions to demonstrate the effectiveness of CaSemCom. This scenario includes both vision-based and text-based tasks across multiple users, allowing us to rigorously evaluate the interplay between the LLM-driven gating mechanism and the multi-expert semantic architecture. The experimental results show that CaSemCom not only accelerates convergence and improves semantic fidelity but also minimizes retransmission overhead compared to basline approaches.}
\end{itemize}


\section{Pillars of Context-Aware Semantic Communication in Wireless Networks} \label{sec:background}

\subsection{Context Awareness in Communications}
From traditional human-to-human interactions to advanced SemCom systems, context always plays a vital role in ensuring communication efficiency.
The context can be categorized into internal and external contexts. 
The former refers to the inherent factors of the entities, e.g., human users, involved in the communication process. 
The latter encompasses environmental conditions, transmission mediums, and situational factors that surround and influence the communication process.

\subsubsection{Linguistic Communications}
Linguistic communication revolves around the interpretation of natural language, where context serves as a cornerstone to align the intention of the sender with the receiver's understanding. In this case, the internal and external contexts are defined as follows.

\begin{itemize}
\item \textbf{Internal Context}: Internal contexts include intrinsic attributes of the communicating entities, such as prior knowledge, cognitive abilities, communication goals, and emotional states. These elements influence how language is generated, interpreted, and aligned with the desired objectives. 
\item \textbf{External Context}: External contexts define the environmental and situational context surrounding communication. These include environmental features, social relationships, dynamic feedback mechanisms, etc. External context sets the conditions under which communication occurs, influencing the strategies (e.g., raising volume in noisy settings, shifting to text-based communication when audio quality deteriorates) employed for linguistic alignment and response generation.
\end{itemize}

The significance of linguistic context in improving communication efficiency is exemplified by the Learning Through Communication (LTC) framework~\cite{wang2024adapting}, which explores the interplay between internal and external contexts in multi-agent systems. Specifically, internal contexts, such as linguistic trajectory\footnote{A liguistic trajectory refers to the complete history of communications and interactions that occurred between an agent and its environment or other agents during a task, including messages exchanged, actions taken, and feedback received.} buffers, enable agents to store and utilize prior contextual knowledge, resulting in a 16.7\% improvement in task completion rates across diverse linguistic scenarios compare to traditional methods without context. External contexts, including task-specific instructions and iterative feedback loops, contribute to a 9.5\% increase in task accuracy by refining the semantic alignment between agents~\cite{wang2024adapting}. 

\begin{figure*}[!t]
    \centering
    \includegraphics[width=0.95\linewidth]{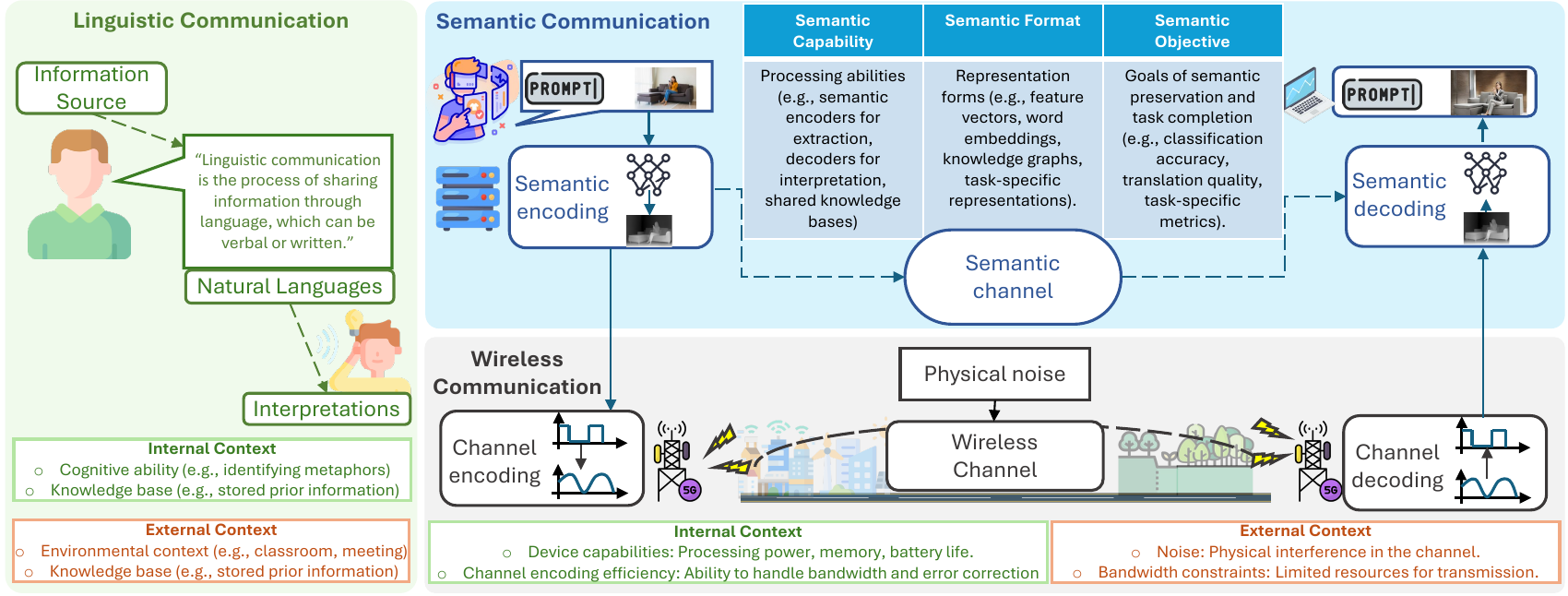}
    \caption{A unified illustration of linguistic communication, SemCom, and wireless transmission, each governed by internal and external contexts. Linguistic communication emphasizes how cognitive abilities, knowledge bases, and situational cues shape the exchange of human language. Wireless communication frames the physical layer, where device constraints, channel fading, and interference further refine which data can be transmitted in real-time. SemCom captures meaning extraction and representation, highlighting how semantic encoders and task objectives drive the transformation of raw data into high-level features. }
    \label{fig:ContextCompare}
\end{figure*}
\subsubsection{Wireless Communications}
As shown in Fig.~\ref{fig:ContextCompare}, wireless communications represent a paradigm shift where information is transferred by bit streams. Correspondingly, the contexts from linguistic communication can be mapped to new technical dimensions \cite{MONTEIRO20191}:
\begin{itemize}
    \item \textbf{Internal Context}: Encompass task-specific contexts that characterize the inherent requirements and capabilities of communicating devices. The requirements define specific performance objectives, such as minimizing latency (e.g., milliseconds for real-time applications), maximizing throughput (e.g., gigabits per second for high-definition video streaming), or achieving target reliability levels (e.g., 99.9\% for ultra-reliable communications). The capabilities describe the device's available resources and constraints, including processing power (e.g., baseband processing capacity), hardware limitations (e.g., number of antenna elements), energy constraints (e.g., battery capacity in mobile devices), and memory capacity for storing and processing communication parameters.
    \item \textbf{External Context}: Define the physical and network environments. Natural aspects include physical transmission mediums (ground, air, or space), working conditions, and weather effects. These manifest in specific challenges. For instance, indoor environments must contend with wall penetration losses and multi-path effects from building structures, while vehicular scenarios have high mobility, causing rapid channel variations and severe Doppler effects. The networking aspects describe the wireless communication states, including channel conditions, signal-to-noise ratios, available bandwidth, and network topology.
\end{itemize}

Context is the foundation for optimizing wireless communication schemes. 
For instance, CarBeam \cite{9662195} captures the road layout and vehicle's location to abstract the vehicular communication environments with high mobility. Additionally, the beam-specific layer-1 reference signal receive power (L1-RSRP) is exploited to depict the wireless channel state. This exploitation of context enables CarBeam to dynamically adjust the time and frequency to perform beam tracking, thereby pursuing the optimal balance between resource costs and accuracy. Compared to context-blind approaches, CarBeam achieves near-optimal throughput with only 1.2-1.4\% degradation from the theoretical optimum while reducing unnecessary beam sweeping overhead by up to 3.5\%.

\subsubsection{Semantic Communications}
An additional layer of context awareness that encompasses multiple semantic dimensions can be integrated into SemCom \cite{LIU2024568}.
The semantic context includes several factors:
\begin{itemize}
    \item \textbf{Semantic Objective}: Defines the goals of semantic preservation and task completion. This includes required semantic accuracy (e.g., classification accuracy for image recognition, translation quality for language tasks), acceptable semantic distortion levels, and task-specific performance metrics.
    \item \textbf{Semantic Format}: Determines how meaning is represented and transmitted in the communication system. This covers various forms like feature vectors for visual content, word embeddings for textual information, knowledge graphs for structured data, or task-specific representations for specialized applications.
    \item \textbf{Semantic Capability}: Encompasses the semantic processing abilities of the communication system, including semantic encoders for meaning extraction, semantic decoders for interpretation, and the knowledge base shared between the sender and the receiver to understand semantic information.
\end{itemize}

As the context becomes a multi-layer framework, SemCom should jointly consider both wireless and semantic aspects. 
For instance, in \cite{10670195}, the authors leveraged Conditional Generative Adversarial Networks-based Channel Estimation (CGE) to estimate channel gains, which indicate how much the transmitted signal is attenuated or amplified by the wireless channel.
With this context information, the decoder can accurately reconstruct the transmitted signal despite channel impairments.
The simulation results demonstrate that the incorporation of wireless context increases transmission accuracy by approximately 5\%. Additionally, a personalized semantic knowledge base is established based on users' individual profiles and task patterns, which enables customized semantic extraction and recovery. Such customized knowledge base settings improve transmission accuracy by around 3\%.


\subsection{Mixture of Expert in Wireless Semantic Communication}
The MoE framework is a dynamic and modular architecture that plays a crucial role in enhancing the efficiency and adaptability of SemCom. By leveraging a specialized set of neural networks working together to complete subtasks (experts) and a gating network, MoE dynamically selects the most relevant experts based on input data and operational context~\cite{he2024toward}. This ensures computational efficiency by activating only the necessary experts while addressing the diverse demands of real-time wireless environments. The ability to adaptively allocate resources makes MoE able to handle the heterogeneous tasks and varying channel conditions characteristic of modern SemCom systems~\cite{shi2021semantic}.

MoE directly aligns with the broader vision of CaSemCom by incorporating both task and communication contexts into its decision-making process. For instance, in vehicular networks, MoE frameworks can mitigate challenges such as semantic noise, secure transmission, and eavesdropping~\cite{he2024toward}. By selectively activating robust experts for noise-resilient encoding, covert experts for secure data transmission, or private experts for enhancing privacy, MoE ensures adaptability across dynamic wireless conditions. Under conditions of source and channel noise, MoE-enabled frameworks maintained communication accuracy 40\% higher than traditional methods~\cite{he2024toward}. In scenarios requiring secure communication, MoE significantly reduced eavesdropping success rates while preserving high semantic fidelity for legitimate receivers.
By integrating MoE into the CaSemCom framework, the system can dynamically adapt to real-time demands while maintaining high performance across different tasks. This integration not only enhances scalability but also aligns with the goal of reducing bandwidth usage and improving semantic fidelity through intelligent, context-driven resource allocation.

{
Recent advancements in communication systems emphasize the growing importance of context-aware mechanisms to address dynamic requirements and environmental challenges. Existing approaches highlight several observations: traditional models often lack the adaptability needed for heterogeneous tasks and conditions; SemCom frameworks remain limited by their static processing pipelines; and integration of task and environmental context is frequently overlooked. Addressing these gaps, the proposed CaSemCom framework leverages multi-expert architectures and context-aware processing to dynamically adapt to real-time demands while optimizing efficiency and semantic fidelity across diverse wireless scenarios and data modalities.}
{
\section{Proposed Context-Aware Semantic Communication Framework}
\label{sec:system_model}
\begin{figure*}[!t]
    \centering
    \includegraphics[width=0.95\linewidth]{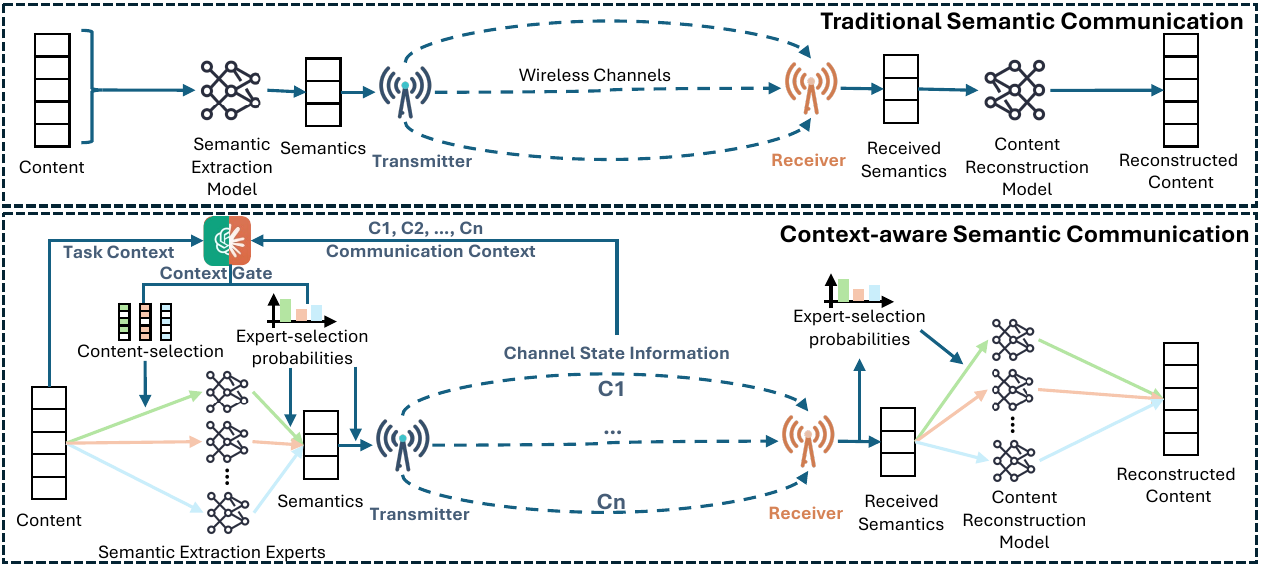}
    \caption{Architecture of the proposed LLM-based CaSemCom framework. At the transmitter, the system determines which parts of the data are semantically important and selects the best expert(s) to encode those parts. The receiver uses the same gating information to ensure accurate semantic reconstruction.}
    \label{fig:framework_architecture}
\end{figure*}
\subsection{Architectural Overview}
Figure~\ref{fig:framework_architecture} presents the core components as follows:
\begin{itemize}
    \item \textbf{Context Aggregator}: At the transmitter, a context aggregator collects both task context (e.g., application priority and latency constraints) and communication context (e.g., channel state information and bandwidth availability). This contextual data includes quantitative parameters and qualitative descriptors, thereby offering a comprehensive view of the operational environment.
    \item \textbf{LLM-Based Gating Mechanism}: Leveraging the contextual information, the LLM-based gating mechanism determines (i) which segments of the input data are most relevant for the application (i.e., input content selection), and (ii) which specialized semantic-extraction model(s) should encode those segments (i.e., expert selection). These gating decisions are optimized based on both user-level requirements and real-time channel conditions, ensuring that only task-critical semantics are prioritized for transmission.
    \item\textbf{Multi-Expert Semantic Architecture}: A pool of specialized encoders (e.g., vision, audio and text) resides at the transmitter, each trained to extract semantic features in a specific modality. The gating mechanism’s decisions activate only the necessary experts, thereby conserving computational and bandwidth resources. A parallel set of decoders at the receiver is likewise coordinated through lightweight control signals that specify how to reconstruct the transmitted semantics.
    \item \textbf{Transmission and Reconstruction}: The transmitter packages semantic features (and minimal gating decisions) for transmission over the wireless channel. Upon receiving input data, the receiver employs the corresponding experts to reconstruct or interpret the data at a semantic level.
\end{itemize}

\subsection{LLM-Based Gating Mechanism}

\begin{figure}[!h]
    \centering
    \includegraphics[width=0.8\linewidth]{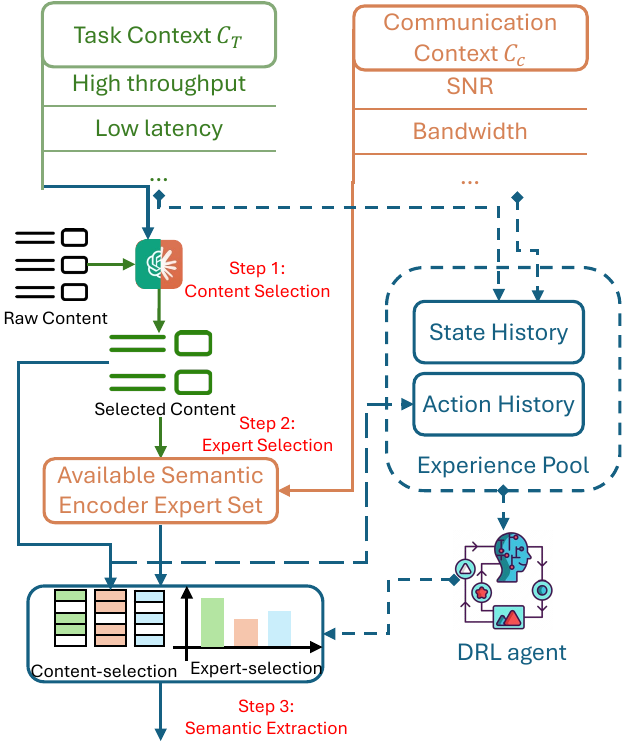}
    \caption{An overview of the LLM-based context gate and its fallback mechanism. Task context ($C_{T}$) and communication context ($C_{c}$) serve as inputs to the LLM, which outputs content-selection and expert-selection decisions. These decisions, alongside the context parameters, are stored in an experience pool as state-action pairs. Over time, a DRL agent is trained on this data and can serve as a fallback decision-maker when LLM inference is unavailable or infeasible.}
    \label{fig:ContextGate}
\end{figure}

The LLM-Based context gate represents the critical intelligence layer in the proposed SemCom framework, enabling nuanced decision-making through the advanced inferential capabilities of LLM. This module analyzes both task-specific and communication contexts to perform adaptive, context-driven optimization of data processing and transmission.

\subsubsection{Input Context Integration}

The LLM-based gate assimilates structured contextual inputs comprising task and communication contexts. Task Context ($C_T$) specifies application-dependent operational parameters, such as latency tolerance, bandwidth requirements, and semantic fidelity thresholds, tailored to diverse applications such as augmented reality or file transfer protocols. Communication Context ($C_c$) encapsulates real-time network performance measures, including signal-to-noise ratio (SNR), channel state information, and bandwidth availability, enabling the system to adapt dynamically to evolving wireless environmental conditions. Integrating these contexts allows the system to balance task-specific requirements against physical-layer constraints.

\subsubsection{Input Content and Expert Selection}

The decision layers within the context gate include input content selection and expert selection, which operate synergistically to optimize SemCom. Input content selection identifies semantically salient portions of the input data, emphasizing task-relevant information and achieving significant reductions in bandwidth utilization. For example, in video conferencing, the context gate prioritizes encoding facial expressions and speech-related audio while suppressing peripheral details. In autonomous vehicular systems, the context gate selectively transmits critical object-detection data, such as bounding boxes around pedestrians or traffic signals.

Accordingly, expert selection dynamically determines a suitable activation of the subset of specialized semantic extraction models based on the selected input content and contextual constraints. The framework incorporates a modular pool of experts, such as text encoders for natural language tasks, audio encoders for speech or acoustic analysis, and vision encoders for image or video feature extraction. For multimodal applications such as teleconferencing, the context gate can concurrently activate multiple experts, ensuring effective utilization of computational resources while satisfying task demands.

{

\textbf{Examples of LLM-based Gating Decisions and Adaptation:}

Consider an AR streaming scenario, initially characterized by a user performing a static AR-assisted repair task. In this context, the LLM-based gating mechanism prioritizes semantic modalities such as detailed depth maps and structural cues (edge detection) to precisely overlay virtual repair instructions onto physical equipment. Consequently, it activates specialized semantic encoders optimized for depth and edge detection to ensure high semantic fidelity.

Now, suppose the context suddenly changes from performing the static repair task to rapidly moving within the environment, transitioning the application from static AR-assistance to real-time navigation guidance. Upon receiving updated task-context information indicating this new dynamic scenario, the CaSemCom framework quickly revises its semantic selection strategy. It dynamically prioritizes human pose estimation to accurately represent user movements and gestures and simplified textual or visual navigation cues.
}

\subsubsection{Fallback Mechanism and DRL Integration}
{
To enhance system robustness, the LLM-based gating networks integrate a fallback mechanism powered by a DRL agent. The fall back mechanism ensures operation continuity when LLM inference is unavailable. A database of state-action pairs (contextual inputs and corresponding gating decisions) generated during LLM inference serves as an experience pool, training the DRL agent to approximate the LLM's decision policies. When LLM inference is unavailable, the DRL agent uses its learned policies to perform expert selection:  
\begin{itemize}
    \item \textbf{State Space:} Combines task context (priority, latency, and semantic fidelity) and communication context (SNR and bandwidth).
\item \textbf{Action Space:} Includes binary decisions on selecting input modalities (edge, pose, segmentation, depth, and text) and activating corresponding semantic experts.
\item \textbf{Reward Function:} Balances semantic accuracy (Image Reward and SSIM) against retransmission overhead, with predetermined weights.

\end{itemize}
Although less sophisticated than LLM-based inference, this mechanism ensures operational continuity with minimal degradation in system performance.
}

\subsection{Multi-Expert Semantic Architecture at the Transmitter}
The multi-expert pool consists of several specialized encoders and decoders, each tuned to handle a specific modality of data or domain of semantic characteristics. Examples include a text-focused encoder for natural language processing tasks, an audio encoder optimized for speech or other acoustic signals, and a vision encoder for images or video segments. The LLM gate's expert selection decisions determine which of these models is activated. By invoking only experts that are truly necessary, the system conserves computational resources and avoids transmitting superfluous data. Each activated expert produces a compressed semantic representation and the associated metadata indicates how the receiver should decode it. In addition to handling different modalities, the framework supports experts within the same modality but trained to extract distinct semantic features. For example, the authors in~\cite{liu2024semantic} proposed the SemCom framework within the vision modality, the experts of which could focus on edge detection, human pose estimation, or image segmentation. These task-specific models allow the system to adapt to diverse application requirements while maintaining efficient resource utilization. By supporting both heterogeneity across modalities and specialization within modalities, the multi-expert architecture enables a versatile and scalable approach to SemCom.

\subsection{Content Reconstruction at the Receiver}
At the receiver, the system interprets the gating signals to activate the corresponding expert decoders. If multiple experts are used, their outputs may be combined through a small fusion module, allowing seamless reconstruction of multimodal data under a unified semantic framework. Depending on the application requirements, the reconstruction process can vary significantly, using different SemCom paradigms.

In cases where reconstruction is not needed, the semantic information itself may suffice for decision-making or interpretation. For example, in real-time object detection tasks for autonomous vehicles, the receiver might use the transmitted bounding-box coordinates directly, bypassing full image reconstruction. This reduces computational overhead and latency while ensuring task-specific precision~\cite{yang2022semantic}.

Alternatively, Generative AI (GenAI) can be utilized in the receiver to reconstruct information while maintaining semantic integrity. This approach is especially useful for tasks that require high-quality content reconstruction under bandwidth constraints. For example, Generative Adversarial Networks (GAN) or diffusion models can generate high-resolution images or detailed 3D models from compressed semantic representations, as demonstrated in applications for immersive virtual reality or remote medical diagnostics~\cite{liu2024semantic}. However, when a full reconstruction of the original data is necessary, the receiver can employ expert decoders to reconstruct the original data with high fidelity. Techniques such as semantic decoding coupled with error correction are applied to mitigate noise introduced during transmission, ensuring minimal semantic loss. 

The receiver periodically reports channel fluctuations, semantic quality, or user experience metrics to the transmitter. This feedback loop enables real-time adaptation of transmission strategies, further optimizing performance in dynamic environments. By employing a flexible and application-specific reconstruction strategy, the receiver ensures that semantic fidelity is preserved across a wide range of use cases, from low-latency applications to high-quality multimedia transmission.

\section{Case Study: Multimodal Vision-Based Semantic Communication in Dynamic Wireless Environments} \label{sec:casestudy}

\subsection{Problem Statement}
We evaluate the effectiveness of the proposed CaSemCom framework in a multimodal, multi-user SemCom scenario aimed at covering a broad range of real-world applications, such as text-guided image generation, Metaverse content delivery, and intelligent edge analytics. 
The following parts elaborate on the dataset composition, semantic modalities, and implementation parameters.
{
\subsubsection{Dataset Construction and Semantic Modalities}
To validate our CaSemCom framework in realistic SemCom scenarios, we constructed a multimodal dataset capturing a broad range of real-world semantic communication conditions. Specifically, the dataset includes:

\begin{itemize}
    \item \textbf{Image-Prompt Categories:} Four diverse categories covering various styles and complexities, including scenery, human-centric poses, objects in indoor settings, and dynamic scenes with multiple entities.
    \item \textbf{Textual Prompts:} One hundred textual prompts generated to span diverse linguistic features and semantic contexts, ensuring meaningful associations with visual scenarios.
    \item \textbf{Tokenization and Embedding:} Each textual prompt was tokenized using the SentenceTransformer model's internal Byte-Pair Encoding (BPE) tokenizer, converting raw text into sub-word tokens. Subsequently, the tokenized text was embedded into consistent 384-dimensional semantic vectors using the SentenceTransformer model (all-MiniLM-L6-v2)\footnote{https://huggingface.co/sentence-transformers/all-MiniLM-L6-v2}. These embeddings capture rich semantic relationships and are used to guide the image generation and annotation processes.
\end{itemize}

For semantic feature extraction, each user's edge device encodes information using five distinct semantic modalities~\cite{liu2024semantic}:

\begin{itemize}
    \item \textbf{Canny Edge Detection:} Extracts contour information for structural analysis.
    \item \textbf{Human Pose Estimation:} Identifies skeletal key points for movement tracking or gesture recognition.
    \item \textbf{Image Segmentation:} Separates foreground objects from the background for scene understanding.
    \item \textbf{Depth Mapping:} Provides a 2.5D representation of the environment, essential for AR/VR applications.
    \item \textbf{Textual Semantics:} Captures linguistic features derived from the embedding process described above.
\end{itemize}

By selectively combining or omitting certain modalities, the framework dynamically adapts to meet specific task requirements (e.g., prioritizing depth mapping for virtual reality) or bandwidth constraints (e.g., excluding pose estimation under poor network conditions).
}

\subsubsection{Implementation Details and Wireless Channel Settings} 
{
We consider a wireless SemCom environment with ten users. An edge device of each user processes raw data (image frames, sensor feeds, and text prompts) into modality-specific feature vectors, which are then transmitted to a generation server that reconstructs images (or interprets textual embeddings) using a diffusion-based model. The final image quality depends on two main factors: (i) the compression level, affected by the available data rate, and (ii) the semantic selection strategy, which is guided by our CaSemCom framework and may prune or downscale certain modalities when bandwidth is limited. All transmissions are subject to Rayleigh fading, with a fixed channel bandwidth of 1.4 MHz. Each user’s signal-to-noise ratio (SNR) varies between approximately $-13\,\mathrm{dB}$ and $30\,\mathrm{dB}$ according to its instantaneous channel gain. This setup reflects realistic operational conditions where users experience heterogeneous channel qualities due to mobility, interference, or varying path losses. The DRL fallback module is trained using a two-layer DQN ($64$ neurons per layer), epsilon-greedy exploration (epsilon decaying from $1.0$ to $0.01$), learning rate $10^{-3}$, discount factor 0.995, replay buffer size $2000$, batch size $32$, over $250$ training episodes.
}
\begin{figure}[tpb]
    \centering
    \includegraphics[width=\linewidth]{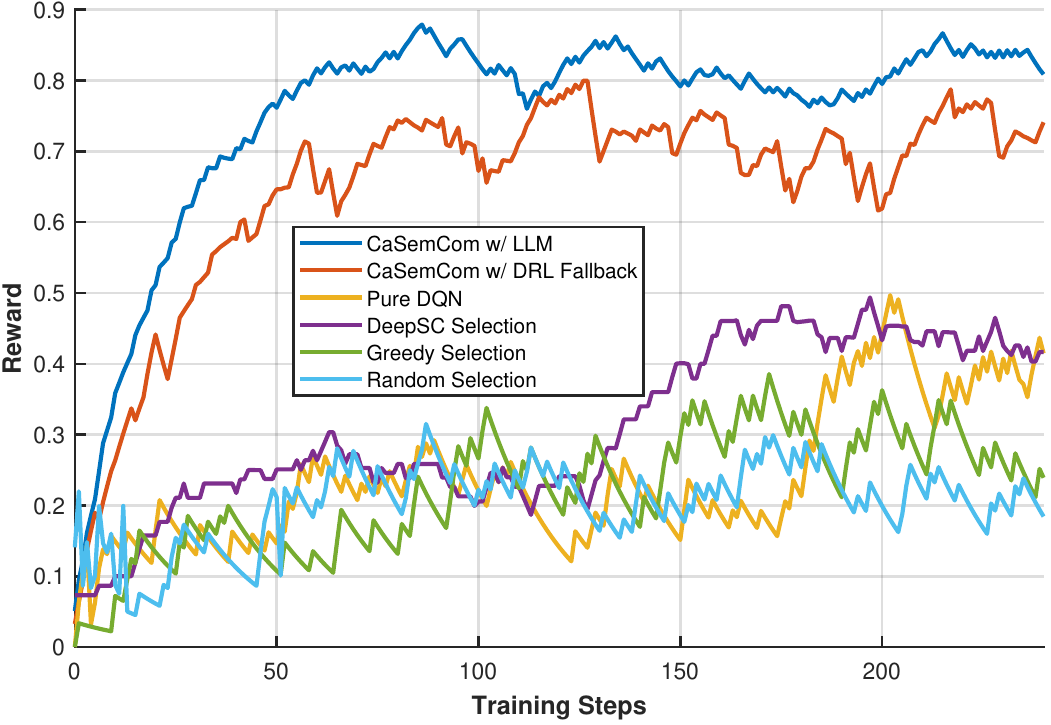}
    \vspace{-0.4cm}
    \caption{Evolution of the average reconstructed image quality, measured by a combined Image Reward and SSIM metric, over 250 training steps. CaSemCom with LLM converges most rapidly and achieves the highest final reward, whereas CaSemCom with DRL fallback converges more gradually but still significantly outperforms the pure DQN, greedy, and random baselines. The DeepSC baseline, despite its learning-based strategy, performs notably worse than CaSemCom with LLM and CaSemCom with DRL fallback, indicating the effectiveness of context-aware gating mechanisms in dynamic semantic communication environments.}
    \label{fig:ex1}
\end{figure}
{
\subsubsection{Performance Indicators}
To quantitatively evaluate and compare different caching strategies, we employed the following performance metrics:
\begin{itemize}
    \item \textbf{Hit Rate:} Defined as the ratio of the number of successfully retrieved items from the cache (cache hits) to the total number of retrieval requests.
    \item \textbf{Caching Overhead:} Quantifies the additional computational or bandwidth resources required due to cache misses or necessary cache updates. Minimizing caching overhead is crucial for efficient resource utilization in bandwidth-constrained wireless networks.
\end{itemize}

Additionally, semantic fidelity was evaluated by comparing reconstructed images against reference images using a combined metric of Image Reward~\cite{xu2024imagereward} (capturing subjective human preference) and Structural Similarity Index Measure (SSIM)~\cite{brunet2011mathematical} (capturing structural similarity). If semantic fidelity fell below a predefined threshold, a retransmission was triggered, incurring further latency and bandwidth overhead.

\subsection{Numerical Results}
\subsubsection{Performance Analysis Over Training Steps} 

Figure~\ref{fig:ex1} illustrates the average reconstructed image quality normalized to [0, 1] over 240 training steps. 
At each step, multiple users are randomly assigned a text or image prompt, and the instantaneous SNR for each user fluctuates according to a Rayleigh fading model. We compare six methods: CaSemCom with LLM, CaSemCom with DRL fallback, pure Deep Q-Network (DQN) \cite{9661349}, Simplified DeepSC~\cite{xie2021deep}, greedy selection, and random selection. 
The reward captures the combined Image Reward and SSIM, reflecting both human preference and structural fidelity.

The two CaSemCom variants demonstrate clear advantages. CaSemCom with LLM converges most rapidly, surpassing a reward value of 0.8 within the first 50 steps. This superior performance results from the LLM interpretation of high-level contextual cues, focusing on key semantic elements even as the channel state varies. Likewise, CaSemCom with DRL fallback converges quickly toward approximately 0.75 because it is specifically designed to handle an LLM outage scenario, simulating conditions in which LLM-based decisions become temporarily unavailable at the probability of 20\%.
In such cases, we utilize the DRL fallback policy trained from past LLM-inferred decisions to handle gating and expert selection. The DeepSC baseline employs a data-driven semantic selection strategy which converges to approximately 0.5, significantly underperforming the two CaSemCom variants, highlighting the effectiveness and necessity of context-aware gating mechanisms as proposed in our CaSemCom framework. On the other hand, the three non-context-aware methods achieve lower rewards. The pure DQN approach converges around 0.4 in approximately 190 steps, underscoring that a single-agent DRL framework without integrated semantic awareness has a limited capacity to adaptively select or compress features in response to dynamic channel variations. The greedy and random schemes yield the worst performance, hovering around 0.3 and exhibiting significant fluctuation. These unstable reward fluctuations are due to incorrect choices, such as prioritizing high-overhead modalities under poor channel conditions, which can severely degrade the reconstructed image quality.
These results confirm that context-aware gating strategies, especially those guided by LLM-based reasoning, are more adept at balancing semantic fidelity with channel constraints. By identifying and transmitting only the most task-critical features, CaSemCom consistently exceeds the baselines in both convergence speed and final quality.

\begin{figure}[tpb]
    \centering
    \includegraphics[width=\linewidth]{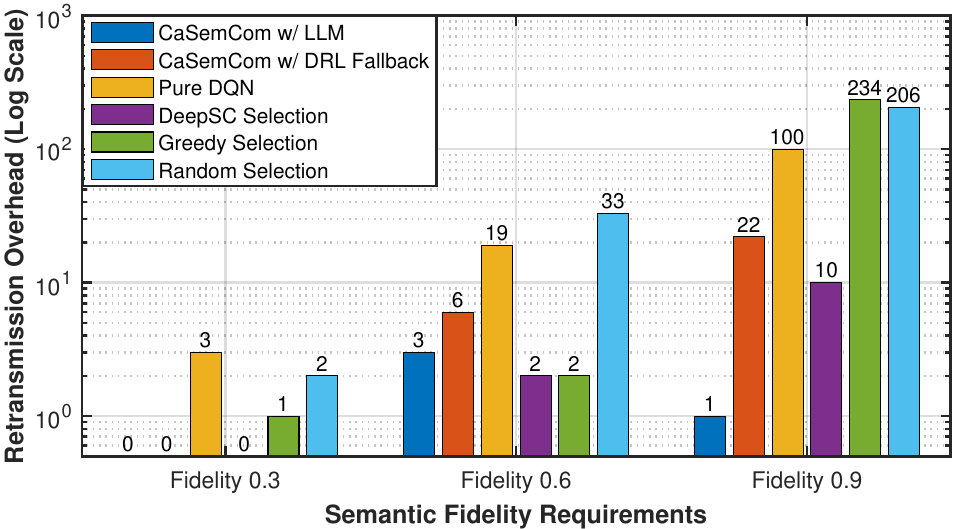}
    \vspace{-0.4cm}
    \caption{Retransmission overhead (log scale) for six methods (CaSemCom with LLM, CaSemCom with DRL fallback, Pure DQN, DeepSC, greedy, and random selection) under different semantic fidelity requirements (0.3, 0.6, and 0.9). CaSemCom with LLM consistently achieves the lowest retransmission overhead, highlighting the advantage of context-aware gating in bandwidth-constrained scenarios.}
    \label{fig:ex2}
\end{figure}
\vspace{0.3cm}

\subsubsection{Retransmission Overhead Under Different Semantic Fidelity Requirements} 
To assess each method's ability to satisfy varying semantic fidelity demands, we set thresholds at 0.3, 0.6, and 0.9. When the transmitted semantics quality fall below the prescribed level, a retransmission is triggered and a unit of retransmission overhead is increased. Figure~\ref{fig:ex2} shows the average retransmission overhead for each method across these thresholds, plotted on a logarithmic scale.
At the lowest threshold of 0.3, overhead remains minimal for all methods, although greedy and random allocations are still marginally worse. As the fidelity requirement increases to 0.6, CaSemCom with LLM emerges as the most resource-efficient, demonstrating substantially fewer retransmissions than any other method. Meanwhile, CaSemCom with DRL fallback exhibits moderate overhead, and pure DQN incurs a considerably higher retransmission frequency. At the most strict threshold of 0.9, the DeepSC baseline incurs notably higher retransmission overhead compared to CaSemCom, while non-learning strategies (greedy and random) also become prohibitively inefficient, requiring repeated transmissions to achieve the required quality. This outcome further emphasizes the limitations of static or insufficiently adaptive semantic selection methods, clearly underscoring CaSemCom's advantage in dynamically and intelligently prioritizing semantic features to efficiently manage retransmissions under stringent fidelity requirements.

An important observation is that although LLM-based context reasoning imposes some overhead in system execution (e.g., model inference time), its ability to prioritize semantic features offers considerable gains at higher-fidelity requirements. In contrast to approaches that indiscriminately send large volumes of data or lack fine-grained semantic awareness, CaSemCom with LLM consistently minimizes repeated transmissions by accurately estimating which features are crucial to meeting the current quality threshold. This capability is especially valuable in practical scenarios where failure to meet critical application-level semantics can render the transmitted data unusable.
}
\section{Conclusion and Future Directions}\label{sec:conclusion}
In this article, we have presented the CaSemCom framework for wireless networks, featuring an LLM-based gating mechanism that dynamically selects both the most relevant input content and the best-suited expert encoder. By incorporating multi-expert semantic extraction and leveraging task- and channel-specific context, CaSemCom simultaneously enhances semantic fidelity and reduces bandwidth consumption, as demonstrated in our multimodal, multi-user case study. Compared to single-agent DRL baselines and non-adaptive methods, CaSemCom converges more rapidly, achieves higher reconstructed image quality, and minimizes retransmissions under high-fidelity requirements.
Despite promising results, several open issues and research directions necessitate further exploration.
\begin{itemize}
    {
\item \textbf{Lightweight Decision Mechanisms:} Although the current DRL fallback mechanism in CaSemCom is designed to be relatively lightweight, it still introduces certain complexity and computational overhead, particularly in resource-constrained or ultra low-latency scenarios. Further research could explore even more lightweight, efficient, yet effective decision-making techniques, potentially through simplified models or hybrid strategies, to enhance performance under stringent computational and latency constraints.
\item \textbf{Adaptive Knowledge Base Management}: The reliance on shared knowledge between the transmitter and receiver to interpret semantic data introduces challenges in large-scale, heterogeneous networks, especially under abrupt context changes. A promising direction is to design adaptive, on-the-fly knowledge synchronization mechanisms that rapidly respond to dynamic user mobility, network conditions, or task requirements. For instance, implementing distributed caching strategies for semantic models or enabling swift, context-specific knowledge updates could significantly enhance CaSemCom's ability to maintain semantic fidelity and responsiveness in rapidly changing communication environments.

    \item \textbf{Enhanced Preprocessing Techniques (Chunking)}: While our current experimental corpus did not require explicit chunking due to the concise nature of each prompt, we recognize that chunking is a crucial preprocessing step for handling larger-scale textual datasets. Developing more efficient and semantically meaningful chunking strategies will significantly enhance the scalability and semantic accuracy of the CaSemCom framework, especially for more complex and longer textual inputs.
  
\item \textbf{Privacy and Security Considerations:} Semantic-level embeddings inherently carry risks of sensitive information leakage. Although CaSemCom mitigates this risk to some extent by dynamically selecting context-dependent semantic modalities which avoided the static use of potentially identifiable semantic representations, there still remains a risk when similar semantics are repeatedly selected over extended periods. A promising future direction to further mitigate this privacy risk is to explore using mixed semantic modalities within a single scene (e.g., combining different semantic embeddings like pose estimation and depth mapping within different spatial regions of an image), making it significantly more difficult for unintended parties to reconstruct sensitive information from semantic embeddings.
  }
\end{itemize}

By addressing these challenges and expanding the framework to additional modalities, knowledge bases, and network architectures, CaSemCom can evolve into an even more versatile and efficient platform for next-generation wireless communication systems.

\bibliographystyle{IEEEtran}
\bibliography{main}
\end{document}